# LEETECH FACILITY AS A FLEXIBLE SOURCE OF LOW ENERGY ELECTRONS

D. Attie[2], S. Barsuk[1], O. Bezshyyko[3], L. Burmistrov[1], A. Chaus[2], P. Colas[2], O. Fedorchuk[3], L. Golinka-Bezshyyko[3], I. Kadenko[3], V. Krylov[1,3], V. Kubytskyi[1], R. Lopez[4], H. Monard[1], V. Rodin[3], M. Titov[2], D. Tomassini[4], A. Variola[1]

[1] *Laboratoire de l'Accélérateur Linéaire (LAL, IN2P3/CNRS and PSud University), Orsay, France*
[2] *Commissariat à l'énergie atomique, Institut de Recherche sur les lois Fondamentales de l'Univers (CEA IRFU), Saclay, France*
[3] *Taras Shevchenko National University of Kyiv (TSNUK), Kyiv, Ukraine*
[4] *CERN, Geneva 23, CH-1211, Switzerland*

A new versatile facility LEETECH for detector R&D, tests and calibration is designed and constructed. It uses electrons produced by the photoinjector PHIL at LAL, Orsay and provides a powerful tool for wide range R&D studies of different detector concepts delivering "mono-chromatic" samples of low energy electrons with adjustable energy and intensity. Among other innovative instrumentation techniques, LEETECH will be used for testing various gaseous tracking detectors and studying new Micromegas/InGrid concept which has very promising characteristics of spatial resolution and can be a good candidate for particle tracking and identification. In this paper the importance and expected characteristics of such facility based on detailed simulation studies are addressed.

## Introduction

Development of new high-energy physics collider experiments calls for a rapid evolution of already established and development of new innovative detector techniques. To characterize new types of detector systems and ensure quality of the already developed instruments high accuracy tests need to be performed. Test beam facilities play a key role in such tests.

Versatile test beam facilities permitting to change beam parameters such as particles type, energy and beam intensity are irreplaceable in characterization and tests of developed instruments. Major applications comprise generic detector R&D, conceptual design and choice of detector technologies, technical design, prototypes and full-scale detector construction and tests, detector calibration and commissioning.

In this article we describe the design and simulation of versatile source of electrons with adjustable bunch intensity and particles energy, LEETECH (Low Energy Electron TECHnique).

The facility is designed to deliver electron samples with energy spread of about 2% and sub-nanosecond time spread within a sample. A LEETECH like facilities are particularly required for a wide range of tests where the energy needs to be changed continuously and detector response to low intensity particle flows (down to 1 particle) needs to be studied. The LEETECH finds its applications in scintillator materials studies for neutrino experiments, studies of particular crystal properties, and quality control for many detector techniques. Particularly relevant are the tests of Micromegas/InGrid technique [1] – the micro-pattern gaseous detector with pixel readout electronics, which provides 3D coordinate measurement and has a performance approaching that of silicon detectors.

The LEETECH source uses electrons from the PHIL facility [2] at LAL, Orsay. The PHIL is a photoinjector and a linear accelerator delivering electron bunches with energy up to 5 MeV with 5 Hz repetition rate and intensity of $10^8$ to $10^{10}$ particles per bunch.

## Existing test beams facilities

A concise overview of existing test beam setups is addressed in this section [3-7]. Test beams are usually obtained from big and expensive accelerator facilities with high energies between tens of MeV and 800 GeV. Operational cost of these test platforms are quite high and user time is deficient. Such an important combination of beam characteristics as small time (tens of picoseconds) and energy (down to 1%) spread, possibility to precisely adjust energy and reduced operational costs is highly required. Precise testing of timing characteristics of modern detectors requires test beams with short bunches in picoseconds range. Therefore LEETECH facility delivering electron samples of low and adjustable energy, variable intensity in the sample, precise timing, accessible and having reduced operational costs, finds its place as a test facility for large variety of detector techniques.

## Photoinjector PHIL

The PHIL (PHotoInjector at LAL) is an electron beam accelerator at LAL. This accelerator is dedicated to tests and characterization of the electron photoguns and high frequency structures for future accelerator projects (lepton colliders of next generation — CLIC, ILC). This machine has been designed to produce electron bunches of low energy (E < 10 MeV), small emittance ($\varepsilon \approx$ 10 $\pi$·mm·mrad), high current (charge nearly 2 nC/bunch) electron bunch at low repetition frequency (<10 Hz) [2]. At the end of the accelerator, the normalized emittance is about 4 $\pi$·mm·mrad. Bunch length is short and determined by laser pulses having FWHM duration of 5 ps. PHIL is currently the 6-meters long accelerator with two extracted beam lines (see Fig. 1).

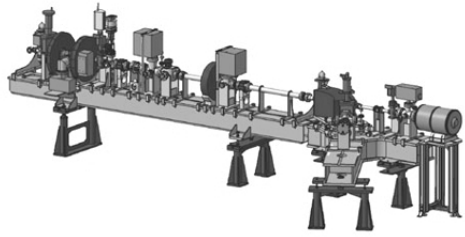

Fig. 1. General view of PHIL photoinjector.

The direct beam line is mainly dedicated to the 2D transverse emittance and bunch length measurement. The deviated line is devoted to the measurements of the beam parameters (mean and dispersion energy measurement of the beam). The injection in the deviated line is performed by the Tesla Test Facility injector dipole.

The direct beam line is equipped with:
- 2 Beam Position Monitors.
- 1 phosphorescent transverse beam profile monitor.
- 1 Faraday Cup.

The beam profile monitor is the phosphorescent screen oriented at 45° relative to the beam axis. The screen is a ceriumdoped yttrium:aluminum:garnet (YAG:Ce) crystal scintillator of 300 μm thickness and 40 mm of diameter.

In 2010, three other phosphorescent YAG:Ce screen monitors were installed on PHIL. The first one is mounted at the entrance of the dipole. It provides important information on the beam behavior just before the dipole magnet, which is used to adjust the beam for the mean and energy spread measurement on the deviate line. Each phosphorescent screen is complemented with a versatile optical system (made of one or more achromatic lens) and a Gigaethernet CCD camera (2 with 1/3" sensor format with 7.4 μm pixel size and 2 with 1/2" sensor format with 4.65 μm pixel size). The CCD dynamic range is 8 bit. In order to avoid pixel saturation during the measurement a remote control optical density wheel is mounted in front of each camera.

One very important feature of the PHIL electron bunches that duration of one pulse is a few of picoseconds and signals from different bunches are very well separated in time which enables us to obtain good time resolution for investigating of detector timing characteristics.

A two meters extension of the direct beam line has been constructed to accommodate the LEETECH facility.

## Principle of the setup

The LEETECH platform uses electrons from PHIL to provide electron samples with adjustable energy and intensity. Schematic view of the setup is shown on Fig. 2:

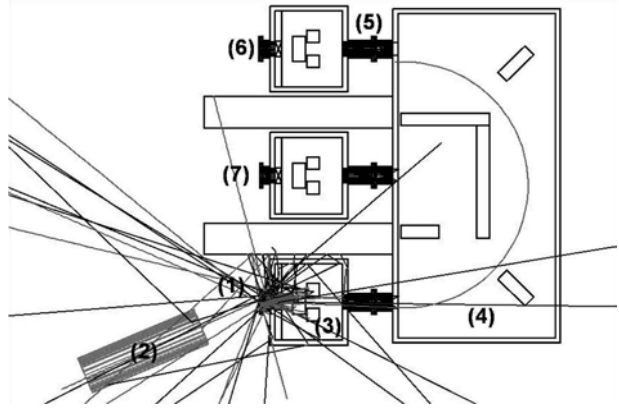

Fig. 2. Geometry implemented into full Geant4 simulation. End of the PHIL beam pipe (2), target (1), vacuum chamber with magnetic field (4), entrance and exit collimator systems (3, 5) and lead shielding are shown.

The principle of setup is described below. An Al attenuator (1) is installed at the exit window of PHIL beam pipe (2). After the attenuator the electrons form a wide energy spectrum and solid angle distribution. The entrance collimators system (3) selects a direction of electrons sent to the spectrometer also adjusting the intensity. Thus obtained narrow secondary beam passes the magnetic field region (4) inside the vacuum chamber. At the exit (5) the electrons are again filtered by exit collimators system and through the thin exit window impinged in the test detector (6). The second exit window (7) with collimators system is dedicated to the training and monitoring purposes.

The electrons of a desired energy are selected by choosing the magnetic field controlled by the current in the dipole magnet. Output intensity can be complementarily adjusted by changing the attenuator thickness and magnetic field. In addition the chosen energy and intensity conditions may define a given choice of the attenuator thickness. Remote control and monitoring of the collimator systems and dipole current is provided via user friendly interface from the PHIL control room.

## Simulation

Before entering the manufacturing phase the full Geant4 simulation of LEETECH platform was performed [8]. Geometry of the setup used in the simulation is shown on Fig. 2. A dedicated shielding in the vacuum chamber suppresses contribution from reflected particles and increases signal to noise ratio.

The full simulation is time and computing-resources consuming, so that 12 hours of simulation time is required using one CPU (Intel Core i5-2410M) to obtain statistical sample of $10^8$ primary particles sufficient to produce a spectrum shown on Fig. 3. To increase the efficiency the MPI (Message Passing Interface) parallel tool [9] was added to the LEETECH simulation under Geant4.9.5 version and later replaced with Multithreading feature for Geant4.10.0 released.

A fake detector volume positioned at the exit of the spectrometer emulates the test detector in the simulation. For all particles entering this volume information about their type, position, momentum and time is stored to the ROOT [10] file. If required this information is used as an input to further simulation. For example this technique is used to simulate the response of diamond detector [12] installed at the exit of the LEETECH spectrometer.

A typical simulation spectrum of electron energy at the exit of LEETECH is shown on Fig. 3.

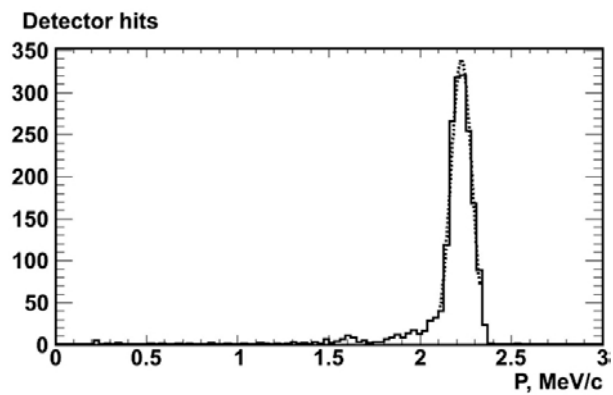

Fig. 3. Typical spectrum of electrons at the exit of
LEETECH facility and Gaussian fit of the peak
(dashed line).

A peak on the spectrum corresponds to the chosen energy value, while the remaining background from scattered particles forms a tail to the left from the signal peak. Simulation is used to optimize shielding and increase signal to noise ratio.

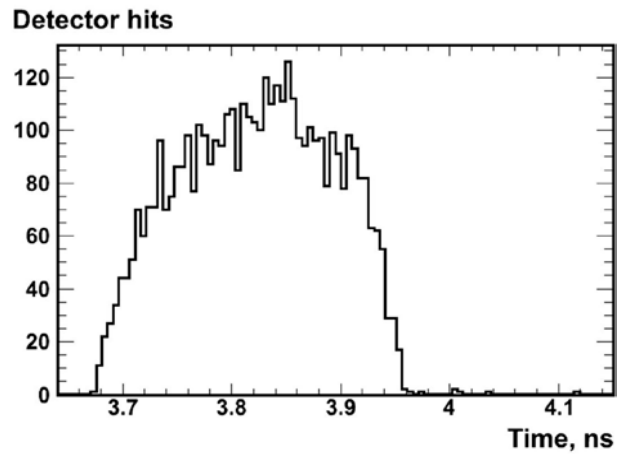

Fig. 4. LEETECH timing characteristics for "wide"
mode of collimator system (simulation).

The timing distribution obtained with the simulation is shown on Fig. 4. This distribution corresponds to the "wide" mode – fully opened collimators (20x20 mm) at the entrance and exit to increase the statistics. Its full width of about 300 ps can still be improved by adjusting the collimators opening.

This full Geant4 simulation is used to validate LEETECH geometry, to study angular and momentum distribution of the electrons after attenuator for different attenuator thickness and to estimate number of delivered electrons as a function of magnetic field, attenuator thickness and collimator opening.

**Attenuator thickness**

Dependence of angular and momentum spectra on the thickness of attenuator was studied. This simulation was performed with the detector positioned behind the attenuator. Momentum spectra of electrons after passing the Aluminum attenuators of different thickness are shown on Fig. 5. The obtained spectra show that different attenuator thickness favours different energy samples, so that attenuators of different thickness are foreseen to be produced.

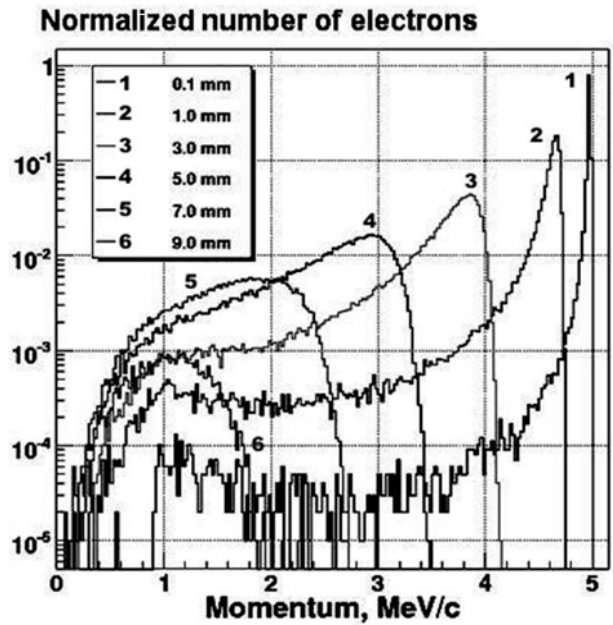

Fig. 5. Momentum spectra of electrons after passing the Aluminum attenuators of different thickness.

The angular distribution of electrons' flight direction after the attenuator is shown on Fig. 6. In agreement to naïve expectations, for thin attenuators the flight direction is smeared around the original flight direction. For thicker attenuators wider distribution of deviation angles is observed. This fact has to also be taken into account for calculations of the intensity of the delivered samples.

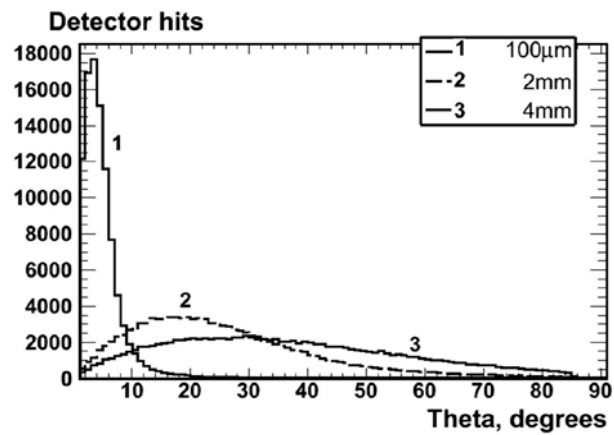

Fig. 6. Angular distribution of electrons' flight direction after the attenuator. Lines numbered 1, 2 and 3 correspond to the 0.1, 2 and 4 mm attenuator thickness, respectively.

**Variation of magnetic field**

In this section we study the intensity of the delivered samples as a function of applied magnetic field. This dependence for different thickness of the attenuator is particularly important since it can be directly measured in the experiment and confronted to the simulation results.

Simulation was performed using primary electrons of 3.5 MeV/c momenta for aluminum targets of 100 μm, 2 mm and 4 mm. The obtained intensities of delivered sample depending on the applied magnetic field are shown on Fig. 7.

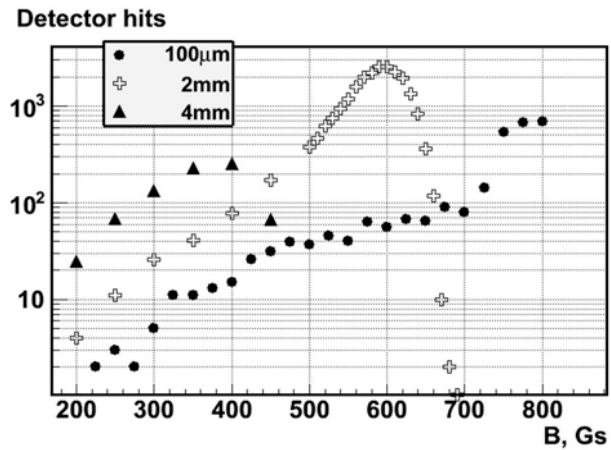

Fig. 7. Magnetic field scans for different attenuator thicknesses. Circles, squares and triangles correspond to the 0.1, 2 and 4 mm attenuator thickness, respectively.

These distributions are in agreement with the spectra shown on Fig. 5. Note that the magnetic field shown on the horizontal axis of Fig. 7 is proportional to the electron momentum shown on the horizontal axis of Fig. 5. For each attenuator, a maximum corresponding to the maximum energy of electrons after the attenuator and slow rising edge, are observed.

However, maximum intensity for the attenuator of 100 μm on Fig. 7 is lower than that for the attenuator of 2 mm. This can be explained by the inclination of the beam pipe with respect to the vacuum chamber by angle of 23º. In case of the 100 μm attenuator the beam solid angle after the attenuator is narrow and the main part of the beam goes aside and hits the collimators. Therefore the number of electrons accessing the vacuum chamber becomes significantly smaller than the total number of electrons after passing the attenuator. In case of the 4 mm attenuator the beam solid angle is wider so bigger part of electrons accesses the chamber.

**Shielding in the vacuum chamber**

The internal shielding of the vacuum chamber was designed to optimize a quality of the delivered samples. Four different geometries of the shielding illustrated by Fig. 8, were suggested and included in the simulation.

The ratio between the number of signal electrons and background hits was chosen as an optimization parameter. The obtained distribution was fit in the signal region with the Gaussian function to estimate the number of signal electrons. Remaining entries were attributed to the background category. The results of this study are shown in Table 1.

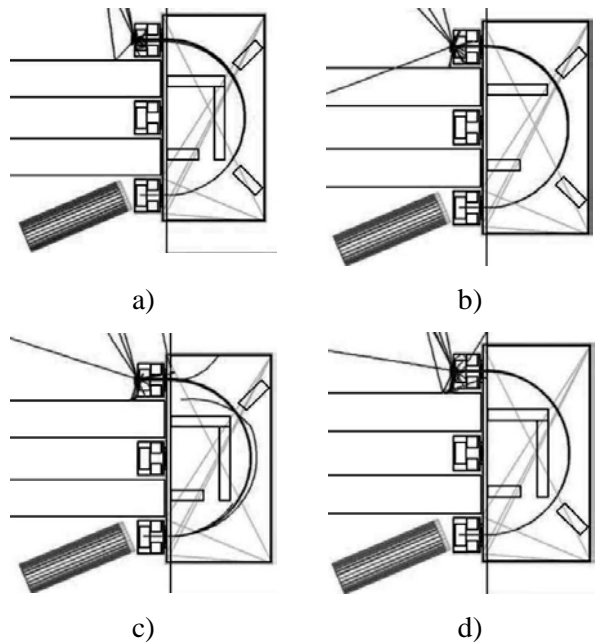

a)  b)

c)  d)

Fig. 8. Different arrangements of the lead blocks
inside the vacuum chamber aiming at the reduction
of contribution from scattered particles.

*Table 1*. Optimization of the shielding in the vacuum chamber using simulation. Number of signal electrons, number of background entries, their ratio and Gaussian width are shown for the geometries outlined on Fig. 8.

| Id | Signal | Background | Ratio, % | σ, keV |
|----|--------|------------|----------|--------|
| A  | 1242   | 139        | 11.2     | 53     |
| B  | 1484   | 242        | 16.3     | 52     |
| C  | 1348   | 175        | 13.0     | 55     |
| D  | 1414   | 186        | 13.2     | 52     |

**Summary**

A new facility LEETECH delivering low-energy electron samples with adjustable energy and intensity of the delivered samples was designed. The R&D, calibration and quality control programs with LEETECH are identified for a wide range of instrumentation techniques. The intensity of delivered samples ranges between few and $10^3$ electrons. Bunches with a duration of hundreds (tens for further studies) of picoseconds are delivered at a frequency of 5 Hz. Samples energy is chosen in the range between hundreds of keV and the maximum PHIL energy of 5 MeV, with resolution of a few percents (down to 1% for further studies). Remote control of the entire setup has been developed.

The full Geant4 simulation of the facility has been performed. Desired parameters of the simulated samples were obtained by adjusting the dipole current and collimators position and by choosing the attenuator thickness. Time of the simulation was significantly reduced using tools for parallel computing. Further improvement of timing characteristics of the delivered samples will be studied.

Next experiments at the built facility are planned to systematically confront simulated parameters of LEETECH to the measured values.

Research was conducted in the scope of the IDEATE International Associated Laboratory (LIA). Work at TSNUK was also partly supported by the State Fund for Fundamental Researches of Ukraine (Research Grant #F58/380-2013, project F58/04 in the frame of the State key laboratory of high energy physics and Research Grant #F69/56-2015).